\documentclass[page-classic]{epl2} 

\usepackage{amsmath,graphicx,tabularx,amsfonts}
\usepackage[pdfpagemode=None,pdfstartview=FitH]{hyperref}
\usepackage{hyperref}
\usepackage{graphicx}
\usepackage{epstopdf}
\usepackage{verbatim}
\usepackage{color}

\DeclareMathOperator{\Tr}{Tr}

\title{Using memory to identify phase transitions on a Cayley Tree}

\author{Javier M. Mag\'an\inst{1} \and Auditya Sharma\inst{1}}
\shortauthor{Javier M. Mag\'an and Auditya Sharma}

\institute{                    
  \inst{1} International Institute of Physics - Federal University of Rio Grande do Norte, Natal, RN, Brazil\\
}

\abstract{We provide a concrete and systematic connection between the statistical physics of the Ising ferromagnet on a Cayley tree, and the study of memory in exponentially expanding spaces. Memory turns out to be a clear signal of the `Bethe-Peierls' phase transition, and the average of memory divided by its standard deviation provides a clear signal of the `spin-glass' transition temperature. Numerical Monte Carlo simulations are used to make transparent the existence of the two different transition temperatures. The quantities used to spot the phase transitions with Monte Carlo could be useful when studying other systems where analytical methods don't work.}

\begin{document}

\maketitle
\section{Introduction}
The objective of this paper is to make a concrete and transparent connection between two separate fields of study that are interested in the same problem~\cite{eggarter1974cayley, matsuda1974infinite, morita1975susceptibility, heimburg1974phase, mullerhartmann1974new, evans2000broadcasting} and to bring to the fore old work, which is of very current interest, in modern language from a contemporary perspective.

Our motivation to revisit an old statistical physics problem came from recent works showing the natural appearance of tree like structures in certain gravity phenomena, like bubble nucleation in eternal inflation~\cite{harlow2012tree} or in the near horizon region of black holes~\cite{barbon2012expanders}. The Ising model in the Cayley tree is complicated enough that some interesting questions that arise here can be meaningfully posed, but still simple enough to answer all of them in a clear way.

In particular, in this article we develop the connection between questions related to memory, which have their root in the branching Markov process approach, and which were raised in a different framework by Ref.\cite{roberts2012memory}, and the more usual static approach, dealing with partition functions and correlators. Building on old knowledge of this system and with the aid of new techniques to compute certain correlation functions in the Cayley tree, we obtain in a clear and simple way the two known critical temperatures, the so called Bethe-Peierls~\cite{eggarter1974cayley} and spin glass~\cite{mezard2001bethe} critical temperatures. Furthermore, we point out that the expectation value of the product of the spin at the root and the total magnetization at the boundary $\langle\sigma_{0}M_{n}\rangle$ (`memory') is the appropriate quantity to signal $T_{BP}$, and 
$\frac{\langle\sigma_{0}M_{n}\rangle}{\Omega}$ where $\Omega = \sqrt{\langle(\sigma_{0}M_{n})^{2}\rangle-\langle\sigma_{0}M_{n}\rangle^{2}}$ is the RMS deviation, signals the spin-glass transition temperature $T_{SG}$. Also, we point out that this conceptual framework can be easily extended to the spin-glass and Heisenberg-type variants of the Hamiltonian - a discussion of some consequences is included.

Monte Carlo simulations on these type of systems in which the boundary is of the same size as the system itself have always been problematic, due to subtleties with the thermodynamic limit. With the help of this new analytic understanding we are able to perform Monte Carlo simulations, showing clearly the two transition temperatures. We expect that the Monte Carlo approach shown here can be used to study other models on the same lattice, where analytical methods fail.

We make clear right at the outset that we consider here the Cayley tree and not the Bethe lattice, the difference between which two has been a source of confusion in the literature - look for example at Ostilli~\cite{ostilli2012cayley}.

\section{The Static Approach: Partition Function}
\begin{figure}
\includegraphics[width=\hsize, trim = 0 145mm 0 0, clip = true]{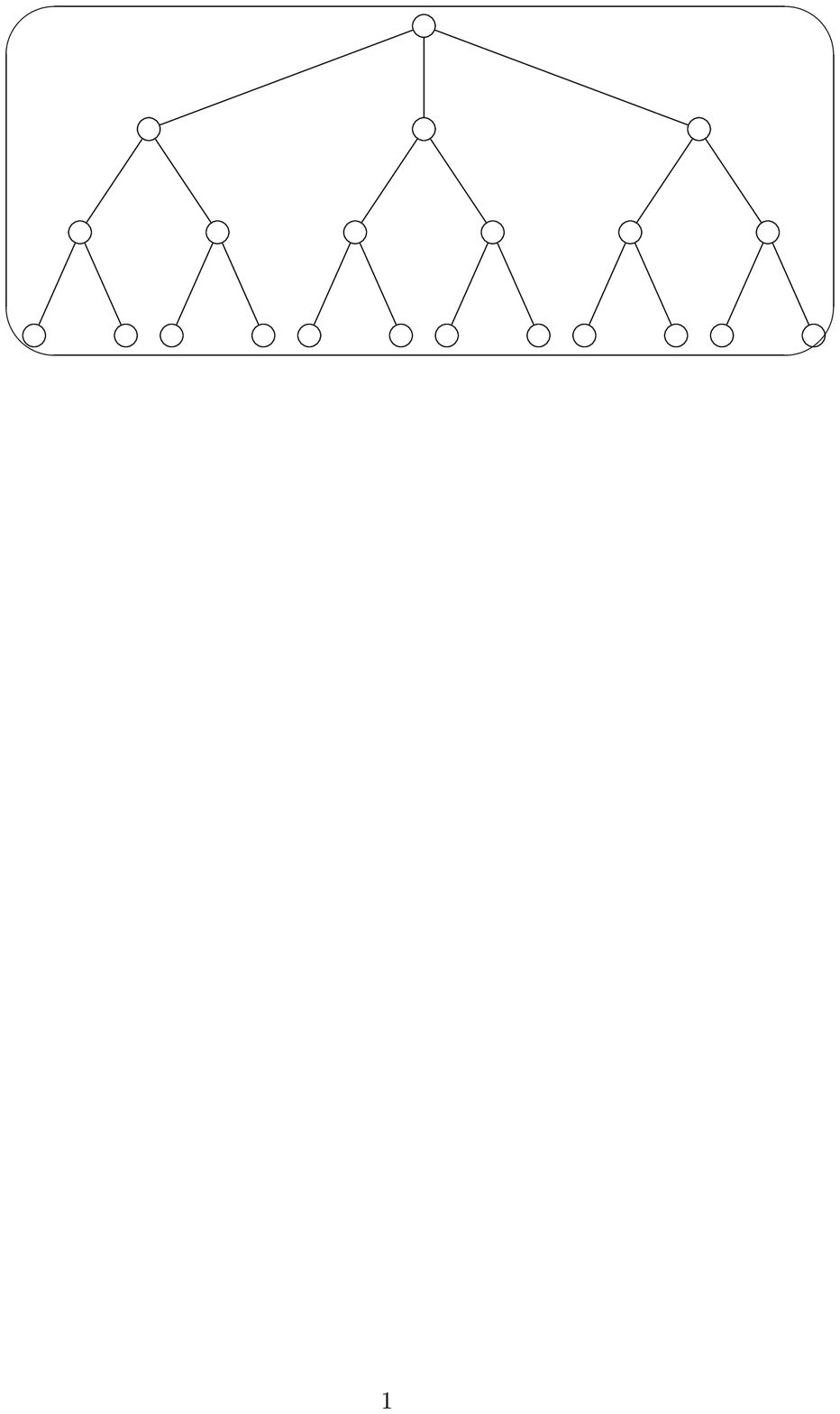}
\caption{Figure shows a Cayley tree of depth $n=3$ and coordination number $z=3$.}
\label{fig1}
\end{figure} 
In 1974, Eggarter~\cite{eggarter1974cayley} provided a beautiful way of computing the partition function of the Ising model on the Cayley tree. The method is based on two important properties. The first one is that, generically, the Ising model Hamiltonian is just a sum of `bonds', i.e, a sum of products of adjacent spins. The second is that the Cayley tree has no loops. Therefore, the bonds can be identified as Ising-like variables in terms of which the Hamiltonian is greatly simplified. More concretely, we start with the following Hamiltonian:
\begin{equation}
\mathcal{H} = -J\sum_{\langle i, j \rangle} \sigma_{i} \sigma_{j},
\label{Ham}
\end{equation}
 where the sum involves pairs of spins which are adjacent on the tree (Fig.~\ref{fig1}). Instead of the spin variables, we choose the spin at the root of the tree $\sigma_{0}$, and the bonds $\theta_{ij} = \sigma_{i}\sigma_{j}$ to define the configuration of our system. The $\theta$ variables can take values $\pm 1$, which make them effectively spin variables too. The Hamiltonian then takes the following simple form
\begin{equation}
\mathcal{H} = -J\sum_{\langle i, j \rangle} \theta_{ij}.
\end{equation}
We see that the Ising model on the Cayley tree is essentially a sum of independent spins (the bonds) subjected to a constant magnetic field; in the computation of the partition function, the only difference with this dual system is a factor of two coming from the sum over the possible states of the spin at the root.  Therefore we arrive at~\cite{eggarter1974cayley}:
\begin{equation}
\mathcal{Z}(J)_{Cayley} = 2(2\cosh(\beta J))^{N_{b}} = 2(2\cosh(\beta J))^{N-1},
\end{equation}
where $N_{b}$ is the number of bonds and $N$ is the number of spins.
With this formalism, namely the $\theta$ variables, the computation of the correlators is straightforward. Firstly, odd correlators vanish due to the spin symmetry of the model.  For the even case we begin with the two point function. We would like to write the spin product in terms of $\theta$ variables. To achieve this, we just need to find the path between the two spins. This path is unique because there are no closed paths in a Cayley tree. Imagine that the path goes over $\sigma_{a},\sigma_{b},\cdots,\sigma_{c}$. We can then write the product as 
\begin{equation}
\sigma_{i}\sigma_{j} = \sigma_{i}\sigma_{a}\sigma_{a}\sigma_{b}\sigma_{b}\cdots\sigma_{c}\sigma_{c}\sigma_{j} = \theta_{ia}\theta_{ab}\cdots\theta_{cj}.
\end{equation}
Now, since the $\theta$ variables are not coupled to each other
\begin{equation}
\langle \sigma_{i}\sigma_{j} \rangle = \langle \theta_{ia}\theta_{ab}\cdots\theta_{cj}\rangle = \langle\theta_{ia}\rangle\langle\theta_{ab}\rangle\cdots\langle\theta_{cj}\rangle = \langle\theta\rangle^{d_{ij}},
\end{equation}
where $d_{ij}$ is the discrete natural distance on the tree, and $\theta$ is a spin variable obeying the following simple Hamiltonian
\begin{equation}
\mathcal{H}_{\theta} = -J\theta
\end{equation}
with partition function
\begin{equation}
\mathfrak{Z} = 2\cosh(\beta J),
\end{equation}
and whose one-point function is given by
\begin{equation}
\langle \theta \rangle = \frac{\sum_{\theta = \pm 1} \theta e^{\beta J \theta}}{\mathfrak{z}} = \frac{e^{\beta J}-e^{-\beta J}}{e^{\beta J}+e^{-\beta J}} = \tanh(\beta J).
\end{equation}
Therefore we have, with the defintion $a \equiv \tanh(\beta J)$ which will prove useful later,
\begin{equation}
\langle \sigma_{i}\sigma_{j}\rangle = \tanh(\beta J)^{d_{ij}} = a^{d_{ij}}.
\end{equation}

Going to higher order correlation functions is also possible. The above technique allows us to obtain the general result for a $2 n$ point function, namely:
\begin{equation}
\langle \sigma_{i_{1}}\sigma_{i_{2}}\cdots \sigma_{i_{2n}}\rangle = (\tanh(\beta J))^{[{\sum_{\langle AB \rangle}d_{i_{A}i_{B}}}]_{min}}
\end{equation}
where $d_{i_{A}i_{B}}$ are the distances between pairs and the minimization is carried out over all possible ways in which $n$-pairs can be formed from $2n$ points. A simple geometric construction to find this minimum is to select $n$-pairs at random, join all the paths between pairs, and delink any portions on the tree that are traversed an even number of times. The newly formed pairs will give a minimum for the sum.\\

\section{The Dynamic Approach: Evolution of a Markov Chain}
Now we consider the type of approach adopted in Refs. \cite{evans2000broadcasting, roberts2012memory}. Consider an Ising spin at time $t=0$, which can be up or down with probabilites $P_{0}=(p_{0}^{\uparrow},p_{0}^{\downarrow})$. At time $t_{1}$ this spin gives rise to $z$ spins, each of which then gives rise to $(z-1)$ spins at the next time instant $t_{2}$, each of which in turn gives rise to $(z-1)$ spins, and so on. There is an obvious tree structure appearing here, with the time $t$ playing the role of the lattice depth in the past static scenario. Now the dynamics is determined by a Markov rule: the probability distribution of a spin at level $t_{n+1}$ is completely characterized by the one of its parent spin at level $t_{n}$. This process defines a Markov chain, in particular a Branching Markov chain, due to the exponential growth of the number of spins at every time step. A Markov chain is defined by its transition matrix, which is obtained here by taking into account that the probability ratios should correspond to the Gibbs distribution.

The joint probability density of the spin at the root and one of its adjacent spins is given by
\begin{eqnarray}
P(\sigma_{0},\sigma_{1}) &= \sum_{\sigma\neq\sigma_{0},\sigma_{1}}\frac{e^{-\beta H}}{Z}=\frac{e^{J\beta\theta_{01}}}{Z}[\Tr_{\theta}e^{-J\theta}]^{N_{b}-1}\\
&=\frac{e^{J\beta \theta_{01}}}{Z}\mathfrak{z}^{N_{b}-1}=\frac{e^{J\beta \theta_{01}}}{2\mathfrak{z}}\nonumber
\end{eqnarray}
and the reduced probability for the spin at the root is just (as expected)
\begin{eqnarray}
P(\sigma_{0}) &= \sum_{\sigma\neq \sigma_{0}}\frac{e^{-\beta H}}{Z} = \frac{\sum_{\theta}e^{-\beta H}}{\sum_{\sigma,\theta}e^{-\beta H}}=\frac{\mathfrak{z}^{N_{b}}}{2\mathfrak{z}^{N_{b}}}=\frac{1}{2}
\end{eqnarray}
The transition matrix is, by definition, the conditional probability $P(\sigma_{1}|\sigma_{0})$, and so
\begin{equation}
G(\sigma_{1},\sigma_{0})=\frac{P(\sigma_{0},\sigma_{1})}{P(\sigma_{0})}=\frac{e^{J\beta \theta_{01}}}{\mathfrak{z}}=\frac{e^{\beta J \sigma_{0} \sigma_{1}}}{\mathfrak{z}}.
\label{eq:transitionmatrix}
\end{equation}
The Markov dynamics is then defined as
\begin{equation}
P_{n+m} = G^{m} P_{n}.
\end{equation}

Now, in order to compute the correlations, we first diagonalize the matrix $G$. With $a \equiv \tanh(\beta J)$,
\begin{equation}
G = \left( \begin{array}{cc}
\frac{1}{\sqrt{2}} & \frac{1}{\sqrt{2}}  \\
\frac{1}{\sqrt{2}} & -\frac{1}{\sqrt{2}} \end{array}\right)  
\left(\begin{array}{cc}
1 & 0  \\
0 & a \end{array}\right) \left( \begin{array}{cc}
\frac{1}{\sqrt{2}} & \frac{1}{\sqrt{2}}  \\
\frac{1}{\sqrt{2}} & -\frac{1}{\sqrt{2}} \end{array}\right)
\end{equation}
and therefore
\begin{equation}
P_{n+m}= \left(\begin{array}{cc}
\frac{1}{2}(1+a^{n}) & \frac{1}{2}(1-a^{n})  \\
\frac{1}{2}(1-a^{n}) & \frac{1}{2}(1+a^{n}) \end{array}\right)   P_{m}
\end{equation}
and as expected, no matter the initial distribution, the probability distribution at late times tends to the stationary distribution $ P=(1/2,1/2) $.

The correlator between any two spins is given by
\begin{equation}
\langle \sigma_{i}\sigma_{j}\rangle = \sum_{\sigma_{i},\sigma_{j}=\pm1} \sigma_{i}\sigma_{j} P(\sigma_{i},\sigma_{j})
\end{equation}
where $ P(\sigma_{i},\sigma_{j}) $ is the probability of finding $ \sigma_{i},\sigma_{j} $. This probability can be found by going to the first common ancestor of the two spins, $\sigma_{r}$. It is given by
\begin{equation}
P(\sigma_{i},\sigma_{j})= \sum_{\sigma_{r}=\pm1} P(\sigma_{i}\vert\sigma_{r})P(\sigma_{j}\vert\sigma_{r})P_{\sigma_{r}}.
\end{equation}
Taking into account that individual spins, and in particular the common ancestor, have stationary distributions in the thermal ensemble, and that the conditional probabilities $ P(\sigma_{i}\vert\sigma_{r}) $ are just given by the elements of the matrix $ G^{d_{ir}} $ we arrive finally at the desired result:
\begin{equation}
\langle \sigma_{i}\sigma_{j}\rangle = a^{d_{ij}}.
\end{equation}
So the Markov chain, which is a time process, gives the same value for the correlators between the spins as the common static procedures on a thermal tree. This observation may be interesting for the model studied in Ref.\cite{harlow2012tree}, which, in this perspective, may be expected to be equal to some statistical model on a thermal tree. Several observations done in that reference become very natural in this perspective (like the p-adic structure of the boundary correlators). We emphasize here that the equivalence between this Markov chain dynamical approach and the static approach is non-trivial; it is not to be confused with the Markov chain that is implicit in a Monte Carlo simulation, which can be performed on, in principle, \emph{any} system, but this equivalence between the dynamical and static approaches is special to this lattice.

\section{The two phase transitions}
\begin{figure*}
(a)\resizebox{0.95\columnwidth}{!}{\input{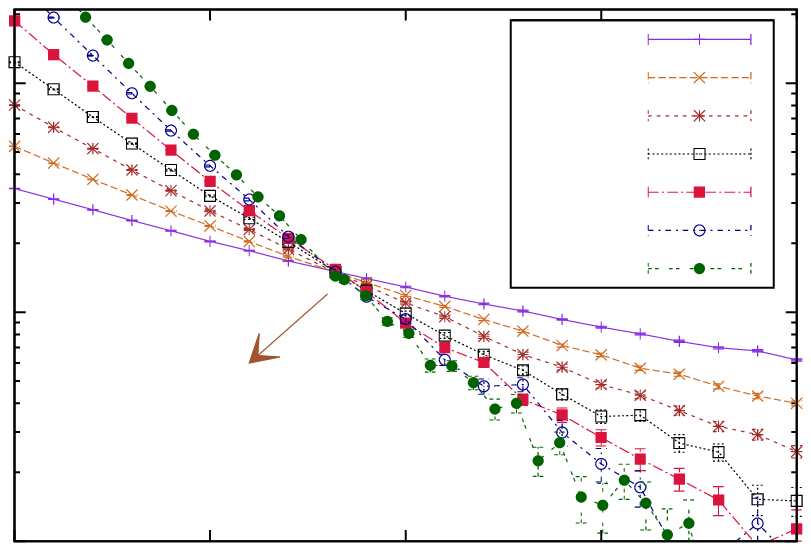}}
(b)\resizebox{0.95\columnwidth}{!}{\input{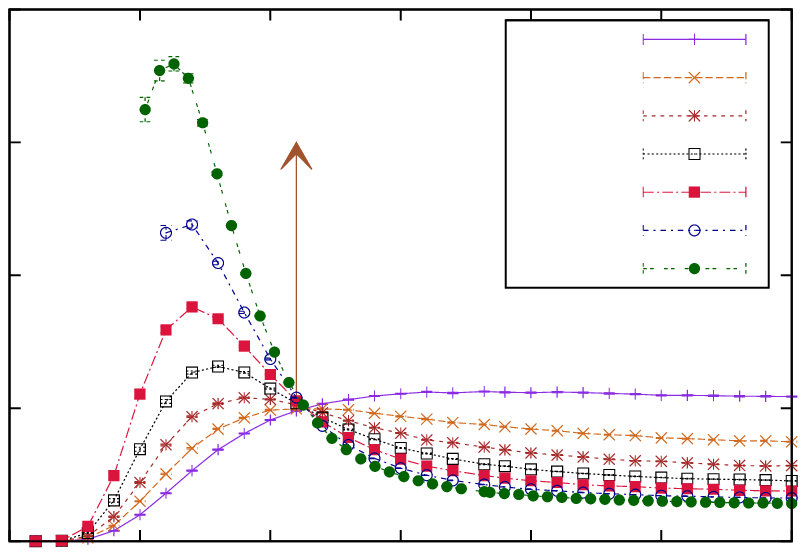}}
\caption{Monte Carlo simulation of the ferromagnet on Cayley trees with coordination number $z=3$. Figure (a) obtains the transition temperature popularly called Bethe-Peierls $T_{BP}$. On the $Y$-axis is plotted the memory, $\langle\sigma_{0}M_{n}\rangle$. Figure (b) obtains the so-called spin-glass transition temperature~\cite{mezard2001bethe}, $T_{SG}$ deeper down in the ordered phase. On the $Y$-axis is plotted the \emph{variance of memory}, scaled by the appropriate finite-size factor.}
\label{fig2}
\end{figure*}
In the Markov chain approach, a very natural question to ask is how correlated some random variable at the $n^{th}$ time-step is with respect to the root spin; in other words, how much \emph{memory} has that random variable about its earliest ancestor. For example, the correlator $\langle \sigma_{0}\sigma_{j}\rangle$ dies for large distances, no matter the value of $ \beta J $. So, as commented in  Ref.\cite{roberts2012memory}, there is no memory for these two random variables. We now show that, in fact, the consideration of the correlation between the \emph{sum} of the spins at the $n^{th}$ time-step with respect to the root-spin, i.e. $\langle\sigma_{0}M_{n}\rangle$, is more useful and signals the so-called `Bethe-Peierls' transition. Furthermore, we will show that the consideration of the allied quantity $\frac{\langle\sigma_{0}M_{n}\rangle}{\Omega}$ where $\Omega = \sqrt{\langle(\sigma_{0}M_{n})^{2}\rangle-\langle\sigma_{0}M_{n}\rangle^{2}}$ is the RMS deviation, 
 gives the so-called spin glass transition temperature. A discussion of the motivation behind studying these quantities is included in the Appendix.

Indeed, that the Ising ferromagnet on a Cayley tree has two distinct phase transitions is long known in the statistical physics community - this information is buried in the papers of Eggarter~\cite{eggarter1974cayley}, Matsuda~\cite{matsuda1974infinite}, and particularly, the papers of J von Heimburg and Thomas~\cite{heimburg1974phase}, and Morita and Horiguchi~\cite{morita1975susceptibility}; however, we believe that our paper shows the correct quantities to study in a MC simulation to obtain these two transitions for the first time.
\subsection{The Bethe-Peierls and the Spin-Glass Transition Temperatures}
Computing the correlation between the root spin and the sum of the spins at the $n^{th}$ level we obtain:
\begin{equation}
\langle \sigma_{0}M_{n}\rangle = \sum_{i\in n^{th} layer} \langle \sigma_{0}\sigma_{i}\rangle = N_{n} \tanh(\beta J)^{n},
\end{equation}
where $ N_{n} $ is equal to the number of spins in the $n^{th}$ layer: $ N_{n}= z(z-1)^{n-1} $, where $z$ is the coordination number for the type of Cayley tree shown in Fig.\ref{fig1}. Therefore, defining $\gamma \equiv z-1$, we have
\begin{equation}
\langle \sigma_{0}M_{n}\rangle = (\gamma \tanh(\beta J))^{n}(1+\frac{1}{\gamma})
\label{eq:memory}
\end{equation}
The behaviour of memory, $\langle \sigma_{0}M_{n}\rangle$ for `long times' or, equivalently, in the `thermodynamic limit' $n\to\infty$, changes completely depending on whether $\gamma \tanh(\beta J)$ is $>$ or $<$ $1$. So, $\gamma\tanh(\beta J)=1$ signals a transition which is understood to be the so-called Bethe-Peierls critical temperature $T_{BP}=\frac{J}{\tanh^{-1}(\frac{1}{\gamma})}$. In fact, $\langle \sigma_{0}M_{n}\rangle$ can show the transition even when $n$ is finite, because at precisely the transition temperature, $(\gamma\tanh(\beta J))^{n}$ becomes independent of $n$, therefore by plotting as a function of temperature for various system-sizes on the same graph, we can obtain the transition temperature. Fig.\ref{fig2}(a) shows that $T_{BP}$ can be obtained from Monte Carlo simulations of finite-systems.

For computing the standard deviation of memory we first need to compute the second order moment. It is given by 
\begin{eqnarray}
\langle(\sigma_{0}M_{n})^{2}\rangle&=\sum_{i,j\in n^{th}layer}\langle\sigma_{i}\sigma_{j}\rangle=\sum_{i,j\in n^{th} layer}a ^{d_{ij}}\nonumber\\
                                   &=(\gamma+1)\gamma^{n-1}+2\sum_{i<j}a ^{d_{ij}}.
\end{eqnarray}
The last sum above was done in Ref.\cite{morita1975susceptibility}. We compute it in a different way with the help of a recursion relation, the details of which are provided in the Appendix.  We collect here the result:
\begin{eqnarray}
\sum_{i<j}a ^{d_{ij}}= N_{n}(\frac{(\gamma-1)}{2\gamma} \frac{x^{n+1}-x}{x-1}+\frac{x^{n}}{2}),
\end{eqnarray}
where $x \equiv \gamma a^{2} \equiv \gamma \tanh^{2}(\beta J)$, and $N_{n} = (\gamma+1)\gamma^{n-1}$ is, once again, the number of spins in the $n^{th}$ layer.
With the help of Eq.\ref{eq:memory}, and some algebra, we can now write down the mean square deviation:
\begin{equation}
\Omega^{2}=\langle(\sigma_{0}M_{n})^{2}\rangle-\langle\sigma_{0}M_{n}\rangle^{2} = N_{n}(1 + \frac{\gamma-1}{\gamma}\frac{x^{n+1}-x}{x-1}-\frac{1}{\gamma}x^{n})
\end{equation}

From this expression one can check that the ratio $\langle\sigma_{0}M_{n}\rangle/\Omega$ defines two phases delimited by the second critical temperature, which is given by $x=1$ ($T_{SG} = \frac{J}{\tanh^{-1}(\frac{1}{\sqrt{\gamma}})}$). In the high temperature limit, $x<1$ this ratio dies in the thermodynamic limit, signalling the absence of memory as discussed previously. In the low temperature regime, $x>1$, the ratio begins to grow, diverging in the strict zero temperature limit. 
For finite $n$, writing
\begin{equation}
\Omega^{2} = N_{n}(1+\frac{\gamma-1}{\gamma}x(1+x+x^{2}+\cdots x^{n-1})-\frac{1}{\gamma}x^{n}),
\end{equation}
we see that when $x=1$, $\Omega^{2} = N_{n}n\frac{\gamma-1}{\gamma}$. Therefore, the quantity $\frac{\Omega^{2}}{N_{n}n}$ becomes independent of system-size at this particular value of the temperature; thus plotting as a function of temperature for various sizes on the same graph, we are able to obtain the spin-glass transition temperature $T_{SG}$. Fig.\ref{fig2}(b) shows $T_{SG}$ obtained from Monte Carlo simulations of a range of finite systems.

\begin{table}[!tb]
\begin{center}
\begin{tabular*}{0.9\columnwidth}{@{\extracolsep{\fill}} c r r r r r r l}
\hline
\hline
$n$ & $N$ & $N_{\rm run}$ & $N_{\rm sweep}$ & $T_{\rm min}$ & $T_{\rm max}$ & $N_{T}$ 
\\
\hline
2 & 10  & 100 &  2048 & 0.10 & 5.00 & 50 \\
3 & 22  & 100 &  2048 & 0.10 & 5.00 & 50 \\
4 & 46  & 100 &  2048 & 0.10 & 5.00 & 50 \\
5 & 94  & 100 &  2048 & 0.10 & 5.00 & 50 \\
6 & 190 & 100 &  4096 & 0.10 & 5.00 & 50 \\
7 & 382 & 100 &  4096 & 0.10 & 5.00 & 50 \\
8 & 766 & 100 &  8192 & 0.30 & 3.00 & 50 \\

\hline
\hline
\end{tabular*}
\caption{
Parameters of the simulations. Here $n$ is the number of layers in the Cayley tree, $N$ is the number of spins (the coordination number was fixed at $z=3$), $N_{\rm run}$ is the number of runs averaged over to extract error-bars, $N_{\rm sweep}$ is the
number of Metropolis Monte Carlo sweeps, $T_{\rm min}$ and $T_{\rm max}$ are the lowest and highest temperatures simulated, and $N_T$ is the
number of temperatures used for parallel-tempering.
\label{tab:fm}}
\end{center}
\end{table}

Our simulations used the standard Metropolis algorithm in conjunction with parallel-tempering which helps speed up the equilibration. Table~\ref{tab:fm} collects the parameters of the simulations. Equilibration was verified by keeping track of the energy; those points (at very low-temperature, and large system sizes) for which the energy did not stay the same within error-bars over the last two sweeps, were discarded.

\subsection{The Spin-glass Transition Temperature from a disordered version of the Hamiltonian}
The reason the second transition temperature is called `spin-glass' is explained in this subsection. Quite remarkably, the methods of Eggarter and others carry right through when one considers a $\pm J$ disordered Hamiltonian on the Cayley tree, and there occurs only one transition temperature at precisely $\frac{J}{\tanh^{-1}(\frac{1}{\sqrt{\gamma}})}$, thus justifying the name `spin-glass' (although see below for a discussion on how there is not a regular type of `spin-glass phase' below $T_{SG}$).

Here we consider the $\pm J$ spin glass on the Cayley tree given by the Hamiltonian
\begin{equation}
\mathcal{H} = \sum_{\langle i, j \rangle} J_{ij} \sigma_{i} \sigma_{j},
\label{Ham}
\end{equation}
where $J_{ij}$ are independent identically distributed random variables that can take values $\pm J$ with equal probability, and where the sum involves pairs of spins which are adjacent on the tree. Instead of the spin variables, like before, we choose the spin at the root of the tree $\sigma_{0}$, and the bonds $\theta_{ij} = \sigma_{i}\sigma_{j}$ to define the configuration of our system. The $\theta$ variables can take values $\pm 1$, which make them effectively spin variables too. The Hamiltonian then takes the following simple form
\begin{equation}
\mathcal{H} = \sum_{\langle i, j \rangle} J_{ij}\theta_{ij}.
\end{equation}
The partition function for this model is identical to that for the Ising ferromagnet since it is an even function in $J$, but more importantly, with the help of the same trick of inserting squares of all the spins that lie between any two given spins, we can now write down the correlator:
\begin{equation}
\langle \sigma_{i}\sigma_{j}\rangle = \pm\tanh(\beta J)^{d_{ij}} = \pm a^{d_{ij}}.
\end{equation}
Whether a plus sign appears or a minus sign appears depends on the number of bonds between $i$ and $j$ that are ferromagnetic. We will now compute an `Edwards-Anderson' type order-parameter except that we consider the correlations between the root spin and the \emph{sum} of the $n^{th}$ layer spins, i.e we are interested in
\begin{equation}
q = [\langle\sigma_{0}M_{n}\rangle^{2}]_{avg},
\end{equation}
where $[\cdots]_{avg}$ means an average over disorder. 

\begin{figure}
\begin{center}
\resizebox{0.95\columnwidth}{!}{\input{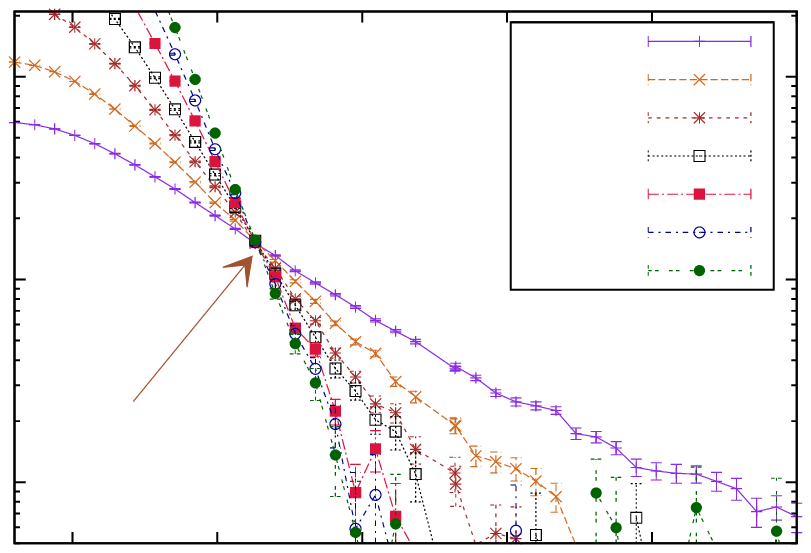}}\
\caption{Monte Carlo simulation of the $\pm J$ spin glass on a Cayley tree with coordination number $z=3$. The figure obtains the transition temperature popularly called Spin-glass $T_{SG}$~\cite{mezard2001bethe}.}
\label{fig3}
\end{center}
\end{figure}

\begin{table}[!tb]
\begin{center}
\begin{tabular*}{0.9\columnwidth}{@{\extracolsep{\fill}} c r r r r r r l}
\hline
\hline
$n$ & $N$ & $N_{\rm samp}$ & $N_{\rm sweep}$ & $T_{\rm min}$ & $T_{\rm max}$ & $N_{T}$ 
\\
\hline
2 & 10  & 100 &  2048 & 0.30 & 3.00 & 40 \\
3 & 22  & 100 &  2048 & 0.30 & 3.00 & 40 \\
4 & 46  & 100 &  2048 & 0.30 & 3.00 & 40 \\
5 & 94  & 100 &  2048 & 0.30 & 3.00 & 40 \\
6 & 190 & 100 &  4096 & 0.30 & 3.00 & 40 \\
7 & 382 & 100 &  4096 & 0.30 & 3.00 & 40 \\
8 & 766 & 100 &  8192 & 0.30 & 3.00 & 40 \\

\hline
\hline
\end{tabular*}
\caption{
Parameters of the simulations on $\pm J$ spin glass. Here, $N_{\rm samp}$ (instead of $N_{\rm run}$) is the number of samples averaged over to extract error-bars, other quantities are as defined in Table~\ref{tab:fm}.
\label{tab:sg}}
\end{center}
\end{table}

Computing the correlation between the root spin and the sum of the spins at the $n^{th}$ level we obtain:
\begin{equation}
\langle \sigma_{0}M_{n}\rangle = \sum_{i\in n^{th} layer} \langle \sigma_{0}\sigma_{i}\rangle = \tanh(\beta J)^{n}\sum_{N_{n} terms} \pm 1,
\end{equation}
where $ N_{n}= z(z-1)^{n-1} $. Squaring we have:
\begin{equation}
\langle \sigma_{0}M_{n}\rangle^{2} = \tanh(\beta J)^{2n}(\sum_{N_{n} terms}\pm 1)(\sum_{N_{n} terms}\pm 1).
\end{equation}
The $\pm 1$ terms in the two sums are completely random, so when we square and average over disorder, only the `diagonal' terms which give $(\pm 1)^{2}=1$ stay, all other terms vanishing because $+1$ is as likely as $-1$.  Therefore, defining $\gamma \equiv z-1$, we have
\begin{equation}
[\langle \sigma_{0}M_{n}\rangle^{2}]_{avg} = (\gamma \tanh(\beta J)^{2})^{n}(1+\frac{1}{\gamma})
\end{equation}
The behaviour of $[\langle \sigma_{0}M_{n}\rangle^{2}]$ for `long times' or, equivalently, in the `thermodynamic limit' $n\to\infty$, changes completely depending on whether $\gamma \tanh(\beta J)^{2}$ is $>$ or $<$ $1$. So, $\gamma\tanh(\beta J)^{2}=1$ signals a transition which is understood to be the so-called spin-glass critical temperature $T_{SG}=\frac{J}{\tanh^{-1}(\frac{1}{\sqrt{\gamma}})}$. Fig.~\ref{fig3} shows that this phase transition can be obtained from Monte Carlo simulations of finite-systems. Again we exploit the convenient fact that at $T_{SG}$, the quantity $(\gamma\tanh(\beta J)^{2})^{n}$ becomes independent of $n$, therefore by plotting $q$ as a function of temperature for various system-sizes on the same graph, we can instantly obtain the transition temperature as the point at which the various curves take the same value. Table~\ref{tab:sg} collects the parameters of the simulations. We make the observation here that the phase below $T_{SG}$ is not really a `spin-glass' in the sense that there is no complex free-energy landscape. Two essential characteristics of a spin glass are frustration and disorder; here, we have only disorder, no frustration. Thus, a `disordered ferromagnet' is perhaps better nomenclature.

\section{Connection to M\"{u}ller-Hartmann and Zittartz, and Extensions}
M\"{u}ller-Hartmann and Zittartz~\cite{mullerhartmann1974new} showed by studying the order at which their free energy expansion develops non-analyticities that the Ising ferromanget on a Cayley tree shows a countable infinity of transition temperatures that they labeled $t_{2},t_{4},t_{8},\cdots,t_{2^{l}},\cdots,t_{\infty}$, with the limit $t_{\infty} = t_{BP}$ being identified as the Bethe-Peierls transition temperaure. In fact, we observe here that their smallest transition temperature $t_{2}$ corresponds to our $T_{SG}$. Since we obtained $T_{BP}$ by considering the mean of memory and $T_{SG}$ by considering the variance of it, it is intriguing to speculate if the other infinite in-between transition temperatures of M\"{u}ller-Hartmann and Zittartz may be harnessed by considering continuous moments of memory that lie between $1$ and $2$.     

Finally, we note that the techniques of Ref.\cite{fisher1964magnetism} can be used in conjunction with the above insights to study the classical Heisenberg ferromagnet on a Cayley Tree. For this case, the function that we have defined as $a=\tanh{\beta J}$ would merely change to $a_{Heisenberg}\equiv\coth{\beta J}-\frac{1}{\beta J}$, and thus the two transition temperatures would be given by $\gamma a_{Heisenberg}=1$, and $\gamma a_{Heisenberg}^{2} = 1$, closely analogous to the Ising case. Indeed the same logic should hold for arbitrary $m$-component vectors where the quantity $a$ is just the ratio of two Bessel functions. Again it would be intriguing to explore such continuous models to understand if an infinite number of transition temperatures are hidden between $T_{BP}$ and $T_{SG}$.

\section{Conclusions and Outlook}
We have shown that the dynamic time Markov chain approach gives exactly the same results as the static ensemble approach of statistical mechanics, thus linking two different communities. We have highlighted that $\langle\sigma_{0}M_{n}\rangle$, which can be thought of as a kind of memory in the time-picture, is a convenient quantity that signals the Bethe-Peierls transition temperature $T_{BP}$. Furthermore, we have shown that $\frac{\langle\sigma_{0}M_{n}\rangle}{\Omega}$, where $\Omega = \sqrt{\langle(\sigma_{0}M_{n})^{2}\rangle-\langle\sigma_{0}M_{n}\rangle^{2}}$ is the RMS deviation, can help identify the `spin-glass' transition temperature $T_{SG}$. We have also shown that a disordered version of the Hamiltonian yields the same transition temperature $T_{SG}$. The analytical understanding allows us to tune our Monte Carlo simulations, which are usually problematic in these kinds of lattices due to the difficulties appearing when taking the thermodynamic limit, to study the right quantities showing clearly the transition temperatures. Ready generalization of our results to arbitrary $m$-component vector models, and some intriguing connections to the infinity of transition temperatures of M\"{u}ller-Hartmann and Zittartz are pointed out. 

The analogue to $\sigma_{0}M_{n}$ in other expanding structures such Euclidean Anti de Sitter or de Sitter space-times is a natural and well-known quantity. Indeed, the fact that in the Cayley tree simple correlators provide the same physical results as the consideration of the more complicated notion of mutual information supports the relation between quantization schemes and the persistence of memory signalled in Ref.\cite{roberts2012memory}. The rate of falloff of the correlators, determined in our case by the temperature and in the EAdS and dS cases by the dimension of the operator being examined, plays the central role in this problem. Exploring the connection better deserves further study.

It would also be interesting to study the correlations between the central node and the sum on the $n^{th} layer$ in quantum models; the entanglement between the central node and the sum on the $n^{th} layer$ could be another intriguing direction to pursue.

\acknowledgments
A.S. thanks Arti Garg, Peter Young for many helpful discussions and for comments on an earlier version of the paper. J.M.M wishes to thank J.L.F. Barb\'on for criticisms and correspondence. We are grateful to Markus M\"uller, Vladimir Gritsev and Dionys Baeriswyl whose observations initiated our discussion on the disordered Hamiltonian. Also both authors gratefully acknowledge the people of Brazil, who, through the CNPQ institution, provide financial support to our research. We are indebted to Achilleas Lazarides for encouragement.

\section{Appendix A }
\label{sec:appA}
Suppose we want to measure how correlated two random variables $A,B$ are. The quantity that is commonly considered in the Markov process community is mutual information, which is defined as
\begin{equation}
I(A,B)=S(A)+S(B)-S(A,B),
\end{equation}
where 
\begin{eqnarray}
S(A)&=-\sum_{A}P(A)\log(P(A)) \nonumber\\
S(B)&=-\sum_{B}P(B)\log (P(B)) \nonumber \quad and\\ 
S(A,B)&=-\sum_{A,B}P(A,B)\log(P(A,B)) 
\end{eqnarray}
are the usual Shannon entropies. This quantity is bounded by zero from below (This happens when $P(A,B)=P(A)P(B)$, which implies that the correlator of the product factorizes $\langle AB\rangle = \langle A\rangle \langle B\rangle$). From above, it is bounded by the entropy of the random variable with lower entropy. This happens when knowing one of them implies knowing the other(for this to happen we need that the entropy stored in the first is at least as much as the entropy stored in the second). For this case, one would expect that the quantity $\langle AB\rangle-\langle A\rangle\langle B\rangle$, which is just the often-studied correlation function in statistical physics, becomes large. However, the \emph{ratio} $\frac{\langle AB\rangle-\langle A\rangle\langle B\rangle}{\sigma_{AB}}$, where $\sigma_{AB}$ is the standard deviation, must also be studied because we may have a scenario where $\sigma_{AB}$ is also large, and thus the knowledge of one of the two random variables $A$ leads to very little predictive power for the variable $B$ even though the two have a large correlation $\langle AB\rangle$. These two different types of quantities help mark precisely the two different phase transitions discussed in the main paper.

\section{Appendix B}
\label{sec:appB}
Here we obtain the sum $\sum_{i<j}a ^{d_{ij}}$, where all $i,j$ are points on the $n^{th}$-layer boundary, with the help of a recursion relation that is obtained by means of a graphical iteration which constructs the tree step by step, from the boundary to the root. We begin by considering two spins at the boundary at distance $2$ in tree units (the minimal distance). At this initial step the sum is, since there are ${\gamma \choose 2}$ ways of choosing the two spins, just trivially ${\gamma \choose 2}a^{2}$ and which provides an initial condition for the iteration. Now we go one step higher, by stitching together $\gamma$ of these $1$-layer trees to one spin above, and thus also connecting spins which are at distance $4$. The sum in this case is found to be $\gamma$ times the first sum, corresponding to the $\gamma$ cases in which the two spins lie within the same old branch, and ${\gamma \choose 2}\gamma^{2}$ terms with value equal to $a^{4}$, which give us the cases in which one spin lies in one old branch and the second spin in another. This pattern is easily seen to hold until one step before the last iteration (the last iteration should include one branch more in the tree we are studying, and has to be treated separately).

Calling this iteration $S_{k}$ the recursion relation takes the following form:
\begin{equation}
S_{k+1}= \gamma S_{k}+{\gamma \choose 2}\gamma^{2k}a^{2(k+1)},
\end{equation}
where $S_{1} = {\gamma \choose 2}a^{2}$.
This can be seen to be a geometric sum by the following transformation $S_{k}\rightarrow {\gamma \choose 2} \gamma^{k-2} C_{k} $. The solution is then
\begin{equation}
S_{k}=  \gamma^{k-2} \frac{x^{k+1}-x}{x-1},
\end{equation}
where $x=\gamma a^{2}$.

Adding $\gamma+1$ braches to form the top root-node (because of the type of Cayley tree we have considered), we have
\begin{equation}
\sum_{i<j}a ^{d_{ij}}= N_{n}(\frac{(\gamma-1)}{2\gamma} \frac{x^{n+1}-x}{x-1}+\frac{x^{n}}{2}),
\end{equation}
where $N_{n}=(\gamma +1)\gamma^{n-1}$ is the number of spins in the last layer. 

\bibliography{ising_cayley}

\end{document}